\documentclass{JHEP3} 

\JHEPspecialurl{http://jhep.sissa.it/JOURNAL/JHEP3.tar.gz}
\usepackage{amsmath,amssymb,epsfig}

\makeatletter
\newcommand{\fmslash}[2][0mu]{%
  \mathchoice
    {\fmsl@sh\displaystyle{#1}{#2}}%
    {\fmsl@sh\textstyle{#1}{#2}}%
    {\fmsl@sh\scriptstyle{#1}{#2}}%
    {\fmsl@sh\scriptscriptstyle{#1}{#2}}}
\newcommand{\fmsl@sh}[3]{%
  \m@th\ooalign{$\hfil#1\mkern#2/\hfil$\crcr$#1#3$}}

\newcommand{\tr}{\hbox{tr}}

\title{\begin{center}
 \bf  \Large   
 Self-energies on deformed spacetimes
 \end{center}}

\author{R. Horvat$^1$, A. Ilakovac$^2$, 
J. Trampeti\'c$^{1,3}$ and J. You$^1$ \\
1. Rudjer Bo\v skovi\' c Institute,
P.O.Box 180, HR-10002 Zagreb, Croatia \\
2. Faculty of Science, University of Zagreb, Bijeni\v{c}ka 32 Zagreb, Croatia\\
3. Max-Planck-Institut f\"ur Physik, (Werner-Heisenberg-Institut),
  	 F\"ohringer Ring 6, D-80805 M\"unchen, Germany\\
E-mail: \email{raul.horvat@irb.hr}, \email{ailakov@rosalind.phy.hr},
\email{josipt@rex.irb.hr }, \email{youjiangyang@gmail.com}}

\received{\today}               
\accepted{\today}               

\abstract{We study one-loop photon ($\Pi$) and neutrino ($\Sigma$) self-energies in a $\rm U(1)$ covariant gauge-theory on d-dimensional noncommutative spaces determined by a antisymmetric-constant tensor $\theta^{\mu\nu}$. For the general fermion-photon ($S_f$) and photon self-interaction ($S_g$) the closed form results reveal self-energies besetting with all kind of pathological terms: the UV divergence, the quadratic UV/IR mixing terms as well as a logarithmic IR divergent term of the type $\ln(\mu^2(\theta p)^2)$. In addition, the photon-loop produces new tensor structures satisfying transversality condition by themselves. We show that the photon self-energy in four-dimensional Euclidean space\-time can be reduced to two finite terms by imposing a specific full rank of $\theta^{\mu\nu}$ and setting deformation parameters $(\kappa_f,\kappa_g)=(0,3)$. In this case the neutrino two-point function vanishes. Thus for a specific point $(0,3)$ in the parameter-space $(\kappa_f,\kappa_g)$, a covariant $\theta$-exact approach is able to produce a divergence-free result for one-loop quantum corrections, having also well-defined both the commutative limit as well as  the pointlike limit of an extended object. While in two-dimensional space  the photon self-energy is finite for arbitrary $(\kappa_f,\kappa_g)$ combinations, the neutrino self-energy still contains an superficial IR divergence.
}


\begin{document}


\section{Introduction}
Following the suggestions from string theory it becomes interesting to study gauge invariant couplings of the known fields together with a certain antisymmetric background tensor $\theta^{\mu\nu}$. Specific subset is found for D-Brane effective action through the observation of the invariance under a deformed gauge symmetry where normal product of fields is replaced by the noncommutative (NC) star($\star$)-product \cite{Seiberg:1999vs}. So one can say that  noncommutative deformation is implemented by replacing the usual dotted product of a pair of fields $\phi(x)$ and $\psi(x)$ by a $\star$-product      $(\phi\star \psi)(x)$ in any action. The specific Moyal-Weyl $\star$-product is relevant for the case of a constant antisymmetric noncommutative deformation tensor $\theta^{\mu\nu}$ and is defined as follows:
\begin{equation}
(\phi\star \psi)(x)=
e^{\frac{i}{2}\theta^{\mu\nu}{{\partial}^\eta_\mu}\,
\,{{\partial}^\xi_\nu}} \phi(x+\eta)\psi(y+\xi)\big|_{\eta,\xi\to0}\;.
\label{f*g}
\end{equation}
Considering coordinates $x^\mu$ as the hermitian operators $\hat x^\mu$ \cite{Jackiw:2001jb}, the coordinate-operator commutation relation is then realized by the following $\star$-commutator of the usual coordinates
\begin{equation}
[\hat x^\mu ,\hat x^\nu]=[x^\mu \stackrel{\star}{,} x^\nu]\equiv
x^\mu \star x^\nu-x^\nu\star x^\mu=i\theta^{\mu\nu}, \;\;
|\theta^{\mu\nu}|\sim\Lambda^{-2}_{\rm NC},
\label{4dim}
\end{equation}
with $\Lambda_{\rm NC}$ being the scale of noncomutativity.
Such a structure arises naturally from geometrical motivations in the branch known as noncommutative geometry \cite{Connes, Madore}. It is reasonable to expect that the new underlying mathematical structures in the NC gauge field theories (NCGFT) could lead to profound observable consequences for the low energy physics. This is realized by the perturbative loop computation first proposed by Filk~\cite{Filk:1996dm}.

There are famous examples of running of the coupling constant in the U(1) NCGFT in the $\star$-product formalism \cite{Martin:1999aq}, and the exhibition of fascinating dynamics due to the celebrated ultraviolet/infrared (UV/IR) phenomenon, without \cite{Minwalla:1999px,Matusis:2000jf}, and with the Seiberg-Witten map (SW) \cite{Schupp:2008fs,Horvat:2011bs,Horvat:2011qg} included. Precisely, in \cite{Minwalla:1999px,Matusis:2000jf} it was shown for the first time how UV short distance effects, considered to be irrelevant, could alter the IR dynamics, thus becoming known as the UV/IR mixing. Some significant progress on UV/IR mixing and related issues has been achieved \cite{Grosse:2004yu,Magnen:2008pd,Blaschke:2009aw,Blaschke:2010ck,Meljanac:2011cs} while a proper understanding of loop corrections is still sought for.

Since commutative local gauge transformations for the D-brane effective action do not commute with $\star$-products, it is important to note that the introduction of $\star$-products induces field operator ordering ambiguities and also breaks ordinary gauge invariance in the naive sense. However both the commutative gauge symmetry and the deformed noncommutative gauge symmetry describe the same physical system, therefore they are expected to be equivalent. This disagreement is remedied by a set of nonlocal and highly nonlinear parameter redefinitions called Seiberg-Witten (SW) map \cite{Seiberg:1999vs}. It appeared then relevant to study ordinary gauge theories with the additional couplings inspired by the SW map/deformation included \cite{Calmet:2001na,Behr:2002wx,Aschieri:2002mc}.

To include a reasonably relevant part of all SW map inspired couplings, one normally calls for an expansion and cut-off procedure, that is, to expansion of the action in powers of $\theta^{\mu\nu}$ \cite{Calmet:2001na,Behr:2002wx,Aschieri:2002mc,Martin:2013gma}. Next follows theoretical studies of one loop quantum properties \cite{Bichl:2001nf,Bichl:2001cq,Grimstrup:2002af,Banerjee:2001un,Martin:2002nr,Buric:2006wm,Latas:2007eu,Buric:2007ix,Martin:2009sg,Martin:2009vg,Buric:2010wd,Buric:2013nja}, as well as studies of some new physical phenomena, like breaking of Landau-Yang theorem, \cite{Schupp:2002up,Minkowski:2003jg,Ohl:2004tn,Alboteanu:2006hh,Alboteanu:2007by,Buric:2007qx}, etc. It was also observed that allowing a deformation-freedom via varying the ratio between individual gauge invariant terms could improve the renormalizability at one loop level~\cite{Buric:2006wm,Latas:2007eu}.

In this work, however, we formulate the $\theta$-exact action-model employing formal powers of fields~\cite{Mehen:2000vs,Jurco:2001my,Okawa:2001mv,Barnich:2003wq,Zeiner:2007,Horvat:2010sr,Horvat:2011iv}, aiming, at the same time, at keeping the nonlocal nature of the modified theory. Introduction of a nonstandard momentum dependent quantity of the type $\sin^2({p\theta k}/2)/(p\theta k/2)^2$ into the loop integrals makes these theories drastically different from their $\theta$-expanded cousins, being thus interesting for pure field theoretical reasons. The deformation-freedom parameters (ratios) are found to be compatible with the $\theta$-exact action therefore included to study their possible effects on divergence cancelation(s).

Two-dimensional noncommutative gauge field theories (2d NCGFT) deserve nonetheless special attention. They are interesting in their own right because of relative simplicity and because in two-dimensions noncommutativity does not break gauge and Lorentz invariance:
\begin{equation}
[x^1 \stackrel{\star}{,} x^2]=i\theta^{12}
=\pm i\Lambda^{-2}_{\rm NC}\;\varepsilon^{12}.
\label{2dim}
\end{equation}
Hence gauge bosons stay massless. Two-dimensional noncommutative Schwinger model \cite{Schwinger:1962tp} was analyzed  recently \cite{Ardalan:2010qb,Armoni:2011pa}. Results at one-loop show that the mass spectrum of commutative and  noncommutative theories is the same.

In this article we obtain a closed forms for fermion-loop and photon-loop corrections to the photon and the neutrino self-energies
using dimensional regularization technique and we combine parameterizations of Schwinger, Feynman, and modified heavy quark effective theory parameterization (HQET) \cite{Grozin:2000cm}. Both two-point functions were obtained as a function of unspecified number of the integration dimensions $D$.
Next we specify gauge field theory dimension $d$ by taking the limits $D\to d$; two different dimensions $d=4,2$ are discussed. 

The paper is structured as follows:
In the following section we describe deformation freedom induced actions, and we give the relevant Feynman rules.  Sections~3 and 4 are devoted to
the computation of photon and neutrino self-energies containing the fermion and the photon loop. Sections~5 and 6 are devoted to discussion and conclusions, respectively. Relevant computational details of the nontrivial loop-integrals are given in the Appendix.

\section{Actions and Feynman rules}

The main principle that we are implementing in the construction of our
$\theta$-exact noncommutative model is that electrically neutral matter fields will be promoted via hybrid SW map deformations \cite{Horvat:2011qn} to noncommutative fields that couple to photons and transform in the adjoint representation of $\rm U_{\star}(1)$. We consider a $\rm U(1)$ gauge theory with a neutral fermion which decouples from the gauge boson in the commutative limit. We specify the action and deformation as a {\it minimal $\theta$-exact completion} of the prior first order in $\theta$ models \cite{Calmet:2001na,Behr:2002wx,Buric:2006wm,Schupp:2002up,Minkowski:2003jg}, i.e. the new (inter-)action has the prior tri-particle vertices as the leading order. So, the minimal gauge invariant nonlocal interaction includes the gauge boson self-coupling as well as  the fermion-gauge boson coupling, denoted here as $S_{\rm g}$ and $S_{\rm f}$, respectively:
\begin{equation}
S=S_{\rm U(1)}+S_{\rm g}+S_{\rm f}.
\label{S}
\end{equation}

Expressing all gauge fields in the action in terms of commutative U(1) field strengths
$f_{\mu\nu}=\partial_\mu a_\nu - \partial_\nu a_\mu$, we obtain the following manifestly gauge invariant expressions:\footnote{In the following we discuss the model construction for the massless case, and set $e=1$. To restore the coupling constant one simply substitutes $a_\mu$ by $ea_\mu$ and then divides the gauge-field term in the Lagrangian by $e^2$. Coupling constant $e$, carries (mass) dimension $(4-d)/2$ in d dimensions.}
\begin{gather}
S_{\rm U(1)}=\int-\frac{1}{4}f_{\mu\nu}f^{\mu\nu}+i\bar\psi\fmslash\partial\psi\,,
\label{U1}\\
S_{\rm g}=\int\theta^{ij}f^{\mu\nu}\left(\frac{\kappa_g}{4}f_{ij}\star_2f_{\mu\nu}-f_{\mu
i}\star_2 f_{\nu j}\right)\,,
\label{g}\\
S_{\rm f}=-\int i\theta^{ij}\bar\psi\gamma^\mu\left(\frac{1}{2}f_{ij}\star_2\partial_\mu\psi-\kappa_f f_{\mu i}\star_2\partial_j\psi\right),
\label{f}
\end{gather}
with the $\star_2$-product being defined as functional $\star$-commutator in \cite{Schupp:2008fs,Horvat:2011bs,Horvat:2011qg}:
\begin{equation}
\phi(x)\star_2 \psi(x)=[\phi(x) \stackrel{\star}{,}\psi(x)]=\frac{\sin\frac{\partial_1\theta
\partial_2}{2}}{\frac{\partial_1\theta
\partial_2}{2}}\phi(x_1)\psi(x_2)\bigg|_{x_1=x_2=x}.
\label{star2}
\end{equation}
Since $S_{\rm g}$ and $S_{\rm f}$ are both gauge invariant by themselves, one can incorporate either or both of them into the full Lagrangian.
The above action were obtained by a $\theta$-exact gauge-invariant truncation of a $\rm U_\star(1)$ model up to tri-leg vertices. Such operation is achievable because the $\rm U(1)$ gauge transformation after deformation preserves the number of fields within each term.

Motivation to introduce deformation parameters $\kappa_g$ and $\kappa_f$ was, besides the general gauge invariance of the action, to help eliminating one-loop pathologies due to the UV and/or IR divergences in both sectors. 
The parameter-space $(\kappa_f,\kappa_g)$ represents a measure of the deformation-freedom in the matter ($S_{\rm f}$) and the gauge ($S_{\rm g}$) sectors, respectively.
Each parameter bears the origin from the corresponding $\theta$-expanded theory too. The gauge deformation $\kappa_g$ was first presented in the non-Abelian gauge sector action of the NCSM and NC SU(N) at first order in $\theta$, $S_{\rm g}^\theta$~\cite{Buric:2006wm,Latas:2007eu}, which could also be realized by modifying the standard SW map for gauge field strength~\cite{Trampetic:2007ze}. We have observed in prior studies \cite{Horvat:2011qg,Horvat:2012vn} that this deformation can be made $\theta$-exact and adopted it here. The deformation-freedom parameter $\kappa_f$ in the photon-gauge boson interaction (\ref{f}) is used to realize the linear superposition of two possible nontrivial NC deformations of a free neutral fermion action proposed in \cite{Horvat:2011qg,Horvat:2011iv}. Its existence was already hinted in the $\theta$-expanded expressions in~\cite{Schupp:2002up} but not fully exploited in the corresponding loop computation yet.

By straightforward reading-out procedure from $S_{\rm g}$ (\ref{g}) we obtain the following Feynman rule for the triple-photon vertex in momentum space:
\begin{equation}
\Gamma_{\kappa_g}^{\mu\nu\rho}(p;k,q)=F(k,q)V_{\kappa_g}^{\mu\nu\rho}(p;k,q);\;\;\;
F(k,q)=\frac{\sin\frac{k\theta q}{2}}{\frac{k\theta q}{2}},
\label{Fg}
\end{equation}
with momenta $p, k, q$ are taken to be incoming satisfying the momentum conservation  $p+k+q=0$. The deformation freedom ambiguity $\kappa_g$ is included in
the vertex function:
\begin{eqnarray}
V_{\kappa_g}^{\mu\nu\rho}(p;k,q)&=&-(p\theta k)\Big[(p-k)^{\rho}g^{\mu\nu}+(k-q)^\mu g^{\nu\rho}+(q-p)^{\nu}g^{\mu\rho}\Big]
\nonumber\\
&-&\theta^{\mu\nu}\Big[p^{\rho}(k q)-k^{\rho}(p q)\Big]-\theta^{\nu\rho}
\Big[k^{\mu}(p q)-q^{\mu}(p k)\Big]
-\theta^{\rho\mu}\Big[q^{\nu}(p k)-p^{\nu}(k q)\Big]
\nonumber\\
&+&(\theta p)^\nu\Big[g^{\mu\rho}q^2-q^\nu q^{\rho}\Big]
+(\theta p)^{\rho}\Big[g^{\mu\nu}k^2-k^\mu k^\nu\Big]+(\theta k)^\mu\Big[g^{\nu\rho}q^2-q^\nu q^{\rho}\Big]
\nonumber\\
&+&(\theta k)^{\rho}\Big[g^{\mu\nu}p^2-p^\mu p^\nu\Big]+(\theta q)^\nu\Big[g^{\mu\rho}p^2-p^\mu p^{\rho}\Big]
+(\theta q)^\mu\Big[g^{\nu\rho}k^2-k^\nu k^{\rho}\Big]
\nonumber\\
&+&\theta^{\mu\sigma}(\kappa_g p+k+q)_\sigma\Big[g^{\nu\rho}(k q)-q^\nu k^{\rho}\Big]
\nonumber\\
&+&\theta^{\nu\sigma}(p+\kappa_g k+q)_\sigma\Big[g^{\mu\rho}(q p)-q^\mu p^{\rho}\Big]
\nonumber\\
&+&\theta^{\rho\sigma}(p+k+\kappa_g q)_\sigma
\Big[g^{\mu\nu}(k p)-k^\mu p^{\nu}\Big] .
\label{Fga}
\end{eqnarray}
The above vertex function (\ref{Fga}) is in accord with corresponding Feynman rule for triple neutral gauge-boson coupling in \cite{Buric:2007qx}.

From $S_{\rm f}$ (\ref{f}) the fermion-photon vertex reads as follows:
\begin{equation}
\Gamma_{\kappa_f}^\mu(k,q)=F(k,q) V_{\kappa_f}^\mu(k,q)
=F(k,q)\Big[\kappa_f\Big(\fmslash{k}(\theta q)^\mu-\gamma^\mu(k\theta q)\Big)-(\theta k)^\mu\fmslash{q}\Big],
\label{Ff}
\end{equation}
where $k$  is the photon incoming momenta, and the fermion momentum $q$ flows through the vertex, as it should.

\section{One-loop photon self-energy}

\subsection{Computing photon two-point function using dimensional regularization}\label{loopD}

Employing the parametrization given in Appendix A and \cite{Horvat:2011qn}, we are enabled to follow the general procedure of dimensional regularization in computing one-loop two point functions. We first present the results with respect to general integration dimension $D$, then in the next sections we will discuss the behavior in different $D\to d$ limits.\\

\begin{figure}
\begin{center}
\includegraphics[width=12cm,height=5cm]{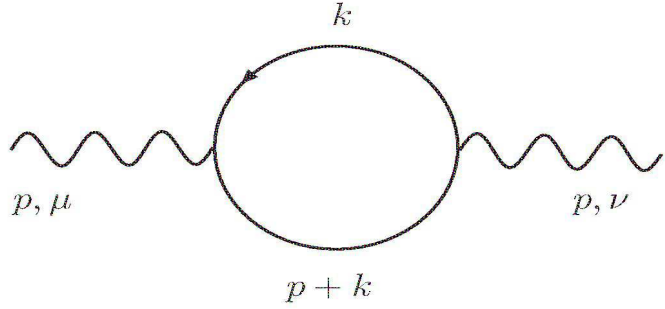}
\end{center}
\caption{Fermion-loop contribution to the photon two-point function}
\label{Fig1}
\end{figure}

\noindent
{\it Photon two-point function: Fermion-loop}\\
The fermion-loop contribution is read out from Fig.\ref{Fig1}
\begin{equation}
\Pi_{\kappa_f}^{\mu\nu}(p)_D=-\tr\;\mu^{d-D}\int\frac{d^D k}{(2\pi)^D}\Gamma_{\kappa_f}^\mu(-p,p+k)\frac{i(\fmslash p+\fmslash k)}{(p+k)^2}\Gamma_{\kappa_f}^\nu(p,k)\frac{i\fmslash k}{k^2}\,,
\label{Fl}
\end{equation}
where the momentum structure and dependence on the parameter $\kappa_f$ is encoded in
\begin{equation}
\tr \;\,V_{\kappa_f}^\mu(-p,p+k)(\fmslash p+\fmslash k) V_{\kappa_f}^\nu(p,k)\fmslash k.
\label{Flmom}
\end{equation}
After considerable amount of computations we have found the following structure:
\begin{equation}
\Pi_{\kappa_f}^{\mu\nu}(p)_D
 =\frac{1}{(4\pi)^2}
 \bigg[\Big(g^{\mu\nu}p^2-p^\mu p^\nu\Big)F_1^{\kappa_f}(p)  + (\theta p)^\mu(\theta p)^\nu F_{2}^{\kappa_f}(p) \bigg],
\label{Flfinal}
\end{equation}
with the loop-coefficients $F_i^{\kappa_f}(p)$
\begin{gather}
\begin{split}
F_1^{\kappa_f}(p) =
 -4{\rm Dim}(Cl[[d]])(4\pi)^{2-\frac{D}{2}}
 &\mu^{d-D}
\; \kappa_f^2\Bigg[\Gamma\Big(2-\frac{D}{2}\Big) \frac{\left(\Gamma(\frac{D}{2})\right)^2}{\Gamma(D)} (p^2)^{\frac{D}{2}-2}
  \\&-2^{\frac{D}{2}-1}\bigg(\frac{(\theta p)^2}{p^2}\bigg)^{1-\frac{D}{4}}\int_0^1 dx \big(x(1-x)\big)^{\frac{D}{4}} K_{\frac{D}{2}-2}\big(X\big)\Bigg],
\end{split}
\label{Ff1}\\
\begin{split}
F_{2}^{\kappa_f}(p)
 ={\rm Dim}(Cl[[d]])(4\pi)^{2-\frac{D}{2}}
 &\mu^{d-D}
 \;\kappa_f\Bigg[\big(\kappa_f-1\big) \bigg(\frac{4}{(\theta p)^2}\bigg)^{\frac{D}{2}} \;\frac{2\Gamma(\frac{D}{2})}{D-1}
 \\&-\kappa_f \;2^{1+\frac{D}{2}}\bigg(\frac{(\theta p)^2}{p^2}\bigg)^{-\frac{D}{4}}
 \int_0^1 dx \big(x(1-x)\big)^\frac{D}{4} K_\frac{D}{2} \big(X\big)\Bigg],
\end{split}
\label{Ff2}
\end{gather}
where ${\rm Dim}\left(Cl[[d]]\right)$ is the dimension of Clifford algebra and $X$ is a new dimensionless variable,
\begin{equation}
X=\sqrt{ x(1-x)p^2(\theta p)^2}.
\label{X}
\end{equation}
The single finite term in (\ref{Ff2}), presenting an additional correction from the SW map induced deformation, vanishes only for $\kappa_f=1$.
All of the divergences arising from the fermion-loop (Fig.\ref{Fig1}) could be removed by the choice $\kappa_f=0$, as in that case the whole general amplitude (\ref{Flfinal}) vanishes for any integration dimensions D.

It is straightforwardly to see that the tensor structure (\ref{Flfinal})
does satisfy the  Ward (Slavnov-Taylor) identity by itself, therefore
\begin{equation}
p_\mu\Pi_{\kappa_f}^{\mu\nu}(p)_D=p_\nu\Pi_{\kappa_f}^{\mu\nu}(p)_D=0.
\nonumber
\end{equation}
\\

\begin{figure}
\begin{center}
\includegraphics[width=12cm,height=5cm]{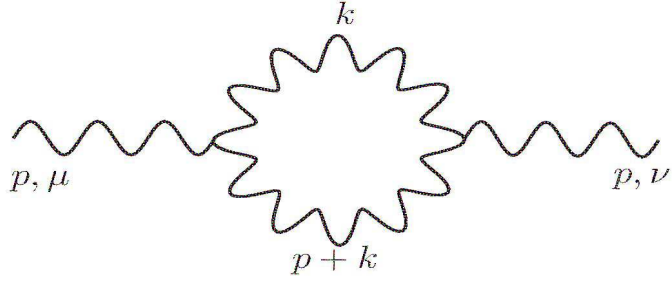}
\end{center}
\caption{Photon-loop contribution to the photon two-point function}
\label{Fig2}
\end{figure}

\noindent
{\it Photon two-point function: Photon-loop}\\
The photon-loop computation involves a single photon-loop integral contribution to photon self-energy from  Fig.\ref{Fig2} in $D$ dimensions
\begin{equation}
\begin{split}
\Pi_{\kappa_g}^{\mu\nu}(p)_D&=
\frac{1}{2}\,\mu^{d-D}\int\frac{d^D k}{(2\pi)^D} \Gamma_{\kappa_g}^{\mu\rho\sigma}
(-p;-k,p+k)\frac{-ig_{\rho\rho'}}{k^2}\Gamma_{\kappa_g}^{\nu\rho'\sigma'}(p;k,-k-p)
\frac{-ig_{\sigma\sigma'}}{(p+k)^2}
\\&=-\frac{1}{2}\,\mu^{d-D}\int\frac{d^D k}{(2\pi)^D} \frac{F^2(p,k)}{k^2(p+k)^2}
V_{\kappa_g}^{\mu\rho\sigma}(-p;-k,p+k)(V_{\kappa_g})^\nu_{\rho\sigma}
(p;k,-k-p).
\end{split}
\label{PldD}
\end{equation}
and as a function of deformation freedom $\kappa_g$ ambiguity correction.
The initial task is to evaluate contractions
$V_{\kappa_g}^{\mu\rho\sigma}(-p;-k,p+k)(V_{\kappa_g})^\nu_{\rho\sigma}(p;k,-k-p)$.

After a lengthy computation we obtained the following compact form of the photon-loop contribution to the photon two-point function in D-dimensions,
\begin{equation}
\begin{split}
\Pi_{\kappa_g}^{\mu\nu}(p)_D&=\frac{1}{(4\pi)^2}\bigg\{\Big[g^{\mu\nu}p^2-p^\mu p^\nu\Big]B_1^{\kappa_g}(p)
+(\theta p)^\mu (\theta p)^\nu B_2^{\kappa_g}(p)
\\&+\Big[g^{\mu\nu}(\theta p)^2-(\theta\theta)^{\mu\nu}p^2
+ p^{\{\mu}(\theta\theta p)^{\nu\}}\Big]B_3^{\kappa_g}(p)
\\&+\Big[(\theta\theta)^{\mu\nu}(\theta p)^2+(\theta\theta p)^\mu(\theta\theta p)^\nu\Big]B_4^{\kappa_g}(p) + (\theta p)^{\{\mu} (\theta\theta\theta p)^{\nu\}} B_5^{\kappa_g}(p)\bigg\}.
\end{split}
\label{PlD}
\end{equation}
Clearly the above structure is much more reacher with respect to earlier $\theta$-exact without SW map results \cite{Hayakawa:1999yt,Brandt:2001ud}. Each of $B^{\kappa_g}_i$ the momentum structures satisfies Ward identities by itself, i.e.
\begin{equation}
p_\mu\Pi_{\kappa_g}^{\mu\nu}(p)_D=p_\nu\Pi_{\kappa_g}^{\mu\nu}(p)_D=0.
\nonumber
\end{equation}
\\

All coefficients $B^{\kappa_g}_i$ can be expressed as sum over integrals over modified Bessel and generalized hypergeometric functions. A complete list of coefficients $F^{\kappa_f}_i(p)$ and $B^{\kappa_g}_i(p)$ as a functions of dimension $D$ is given in the Appendix B.

\subsection{Photon self-energy at different dimensions}

\noindent
{\it Fermion loop in the $D\to4$ limit}\\
In the limit $D\to 4-\epsilon$, the loop-coefficients can be expressed in the following closed forms:
\begin{equation}
\begin{split}
F_1^{\kappa_f}(p) =& -\kappa_f^2\;\frac{8}{3}\Bigg[\frac{2}{\epsilon}+\ln{\pi e^{\gamma_E}}+\ln\big(\mu^2(\theta p)^2\big)\Bigg]
\\&+4\kappa_f^2p^2(\theta p)^2\sum\limits_{k=0}^\infty\frac{(k+2)(p^2(\theta p)^2)^k}{4^k\Gamma[2k+6]}
\\&\cdot
\bigg[(k+2)\Big(\ln \big(p^2(\theta p)^2\big)-\psi(2k+6)-\ln4\Big)+2\bigg],
\end{split}
\label{FlfinalF1}
\end{equation}
\begin{equation}
\begin{split}
F_{2}^{\kappa_f}(p) =& \kappa_f\;\frac{8}{3}\frac{p^2}{(\theta p)^2}\Bigg[\kappa_f-8\big(\kappa_f+2\big)\frac{1}{p^2(\theta p)^2}\Bigg]
\\&-4\kappa_f^2p^4\sum\limits_{k=0}^\infty\frac{(p^2(\theta p)^2)^k}{4^k\Gamma[2k+6]}
\\&\cdot
\bigg[(k+1)(k+2)\Big(\ln\big(p^2(\theta p)^2\big)-2\psi(2k+6)-\ln4\Big)+2k+3\bigg].
\end{split}
\label{FlfinalF2}
\end{equation}
The above expressions for $F_{1,2}^{\kappa_f}(p)$ contain both contributions, from the planar as well as from the non-planar graphs.
\\

\noindent
{\it Photon loop in the $D\to 4$ limit}\\
Using each $B^{\kappa_g}_i(p)$ in (\ref{PlD}) from the Appendix B in the $D\to 4-\epsilon$ limit, we found expressions similar to the fermion-loop. Next we concentrate on the {\em divergent} parts in the IR regime
\begin{equation}
\label{PIB1}
\begin{split}
B_1^{\kappa_g}(p)\sim&\bigg(\frac{2}{3}\big(\kappa_g-3\big)^2 +\frac{2}{3}\big(\kappa_g+2\big)^2\; \frac{p^2(\tr\theta\theta)}{(\theta p)^2}
+\frac{4}{3}\big(\kappa^2_g+4\kappa_g+1\big)\; \frac{p^2(\theta\theta p)^2}{(\theta p)^4}\bigg)
\\&\cdot\left[\frac{2}{\epsilon} + \ln(\mu^2(\theta p)^2)\right]
\hspace{3mm}-\frac{16}{3}\big(\kappa_g-1\big)^2\;\frac{1}{(\theta p)^6}\bigg((\tr\theta\theta)(\theta p)^2+4(\theta\theta p)^2\bigg)\,,
\end{split}
\end{equation}
\begin{equation}
\label{PIB2}
\begin{split}
B_2^{\kappa_g}(p)\sim&\bigg(\frac{8}{3}\big(\kappa_g-1\big)^2\; \frac{p^4(\theta\theta p)^2}{(\theta p)^6}+\frac{2}{3}\big(\kappa_g^2-2\kappa_g-5\big)\frac{p^4(\tr\theta\theta)}{(\theta p)^4}
+\frac{2}{3}\big(25\kappa^2_g
\\&-86\kappa_g+73\big)\frac{p^2}{(\theta p)^2}\bigg)\left[\frac{2}{\epsilon} + \ln(\mu^2(\theta p)^2)\right]
-\frac{16}{3}\big(\kappa_g-3\big)\big(3\kappa_g-1\big)\frac{1}{(\theta p)^4}
\\&+\frac{32}{3}(\kappa_g-1\big)^2\frac{1}{(\theta p)^8}\bigg((\tr\theta\theta)(\theta p)^2+6(\theta\theta p)^2\bigg),
\end{split}
\end{equation}
\begin{gather}
B_3^{\kappa_g}(p)\sim-\frac{1}{3}\big(\kappa_g^2-2\kappa_g-11\big)\frac{p^2}{(\theta p)^2}
\left[\frac{2}{\epsilon} + \ln(\mu^2(\theta p)^2)\right]
-\frac{8}{3(\theta p)^4}\big(\kappa_g^2-10\kappa_g+17\big),
\label{PIB3}
\\
B_4^{\kappa_g}(p)\sim-2\big(\kappa_g+1\big)^2\frac{p^4}{(\theta p)^4}\left[\frac{2}{\epsilon} + \ln(\mu^2(\theta p)^2)\right]-\frac{32p^2}{3(\theta p)^6}\big(\kappa_g^2-6\kappa_g+7\big),
\label{PIB4}
\\
B_5^{\kappa_g}(p)\sim\frac{4}{3}\big(\kappa_g^2+\kappa_g+4\big)\frac{p^4}{(\theta p)^4}
\left[\frac{2}{\epsilon} + \ln(\mu^2(\theta p)^2)\right]+\frac{64p^2}{3(\theta p)^6}\big(\kappa_g-1\big)\big(\kappa_g-2\big).
\label{PIB5}
\end{gather}
Note that all $B_i^{\kappa_g}(p)$ coefficients are computed for abitrary $\kappa_g$ and the notation $\sim$ means that in the above equations we have neglected all finite terms. We observe here the presence of the UV divergences as well as quadratic UV/IR mixing in all $B^{\kappa_g}_i$'s. The logarithmic IR divergences from planar and nonplanar sources appear to have identical coefficient and combine into a single $\ln\mu^2(\theta p)^2$ term. Finally no single $\kappa_g$ value is capable of removing all novel divergences.
\\

\noindent
{\it Fermion and photon loops in the $D\to2$ limit}\\
By setting the $D\to 2-\epsilon$ instead of the $D\to 4-\epsilon$ limit in (3.4) and (3.5), we are able to  compare our results with the results of \cite{Armoni:2011pa}, obtained in the U(N) model of 2d NCGFT.

One could easily observe that $F_{1,2}^{\kappa_f}(p)$ remain finite when $D\to d=2$, therefore we can directly set $D=d=2$.
One can then see that the Bessel K-functions in $F_1^{\kappa_f}(p)$ and
$F_{2}^{\kappa_f}(p)$ for arbitrary $\kappa_f$ exactly cancels each other since $K_{-1}[z]=K_1[z]$. Then, as an example for $\kappa_f=1$, the dimension of Clifford algebra ${\rm Dim}\left(Cl[[2]]\right)=2$ yields:
\begin{gather}
\begin{split}
F_1^{\kappa_f=1}(p) =
 -32\pi
 \Bigg[(p^2)^{-1}-\bigg(\frac{(\theta p)^2}{p^2}\bigg)^{\frac{1}{2}}\int_0^1 dx \big(x(1-x)\big)^{\frac{1}{2}} K_{-1}\big(X\big)\Bigg],
\end{split}
\label{F12}\\
\begin{split}
F_{2}^{\kappa_f=1}(p)
 =-32\pi\bigg(\frac{(\theta p)^2}{p^2}\bigg)^{-\frac{1}{2}}
 \int_0^1 dx \big(x(1-x)\big)^{\frac{1}{2}} K_1 \big(X\big).
\end{split}
\label{F22}
\end{gather}

In two dimensions the totally antisymmetric tensor is up to a constant factor unique and rotationally invariant, so the tensor structures appearing in (\ref{Flfinal}) coincide
\begin{equation}
(\theta p)^\mu(\theta p)^\nu p^2=\big(g^{\mu\nu}p^2-p^\mu p^\nu\big)(\theta p)^2,
\label{2ts}
\end{equation}
therefore
\begin{equation}
\Pi_{\kappa_f}^{\mu\nu}(p)_2=\frac{1}{(4\pi)^2}\Big(g^{\mu\nu}p^2-p^\mu p^\nu\Big)\bigg[F^{\kappa_f}_1(p)+\frac{(\theta p)^2}{p^2}F^{\kappa_f}_2(p)\bigg],
\label{Pi2}
\end{equation}
and we are left with a single finite term from the planar integral.
Summing the two terms in \eqref{Pi2} makes the modified Bessel function integrals from (\ref{F12}) and (\ref{F22}) to cancel each other for arbitrary $\kappa_f$, representing in fact a cancellation of non-planar graphs, and leaving the  total amplitude in the following form
\begin{equation}
\Pi_{\kappa_f}^{\mu\nu}(p)_2
 =\frac{1}{4\pi}\Big(g^{\mu\nu}p^2-p^\mu p^\nu\Big)\bigg[
 \frac{8}{p^2}\kappa_f\big(\kappa_f-2\big) \bigg].
 \label{pkf}
\end{equation}

In the limit $\kappa_f \to 0$, we clearly have $g_{\mu\nu}\Pi_{\kappa_f=0}^{\mu\nu}(p)_2=0$.
Taking the limit $\kappa_f \to 1$, and restoring the coupling constant $e$, we have found photon self-energy to be finite:
\begin{equation}
g_{\mu\nu}\Pi_{\kappa_f=1}^{\mu\nu}(p)_2\,
  =-\frac{2e^2}{\pi}\,.
\label{F2ng}
\end{equation}
Thus the Eq. (15) in \cite{Armoni:2011pa} still holds. Compare our planar result with Eq. (12) in \cite{Armoni:2011pa} revolves also an extra factor $-2$. Noticing the elimination of phase factor $i$ from Eq. (8) to (11) in \cite{Armoni:2011pa}, we identify the minus sign as the effect of Wick rotation performed there.
The missing of factor two could be due to a certain differences regarding the definition of the starting action.

Analyzing photon-loop contribution to the photon self-energy in two Euclidean dimensions (d=D=2), appart from (\ref{2ts}) we have to use the following simplifications $(\theta\theta)^{\mu\nu}=-\Lambda_{\rm NC}^{-4}g^{\mu\nu}=-\Lambda_{\rm NC}^{-4}\delta^{\mu\nu}$, $(\theta\theta p)^{\mu}=-\Lambda_{\rm NC}^{-4}p^{\mu}$, $(\theta\theta\theta p)^{\mu}=-\Lambda_{\rm NC}^{-4}(\theta p)^{\mu}$, $(\theta p)^2=\Lambda_{\rm NC}^{-4}p^2$, and $(\theta\theta p)^2=\Lambda_{\rm NC}^{-8}p^2$. We further fix $\tr\theta\theta=-D\Lambda_{\rm NC}^{-4}$ for any nondegenerate $\theta^{\mu\nu}$ that satisfies $(\theta\theta)^{\mu\nu}=-\Lambda_{\rm NC}^{-4}g^{\mu\nu}=-\Lambda_{\rm NC}^{-4}\delta^{\mu\nu}$ according to the dimensional regularization prescription. Thus the photon-loop contribution can be reduced to one term too. After restoring the coupling constant $e$ we obtain the following contributions from the photon-loop
\begin{equation}
\begin{split}
\Pi_{\kappa_g}^{\mu\nu}(p)_2&=\frac{e^2}{(4\pi)^2}\Big[g^{\mu\nu}p^2-p^\mu p^\nu\Big]\bigg(B^{\kappa_g}_1+\frac{B^{\kappa_g}_2+2B^{\kappa_g}_3}{\Lambda_{\rm NC}^4}-\frac{B^{\kappa_g}_4+2B^{\kappa_g}_5}{\Lambda_{\rm NC}^8}\bigg)
\\&=\frac{e^2}{4\pi}\Big[g^{\mu\nu}p^2-p^\mu p^\nu\Big] B^{\kappa_g}(p).
\end{split}
\label{Pl2}
\end{equation}
Explicit evaluation of $B^{\kappa_g}(p)$ using the integrals defined in Appendix B yields the following finite photon-loop result
\begin{equation}
B^{\kappa_g}(p)=\frac{16}{p^2}\Big(\kappa_g^2-7\kappa_g+7\Big).
\label{B}
\end{equation}
Two $\kappa_g$ values $(7\pm\sqrt{21})/2$ gives vanishing $B^{\kappa_g}(p)$.


\subsection{Photon loop with a special $\theta^{\mu\nu}$ in the $D\to 4$ limit}

In our prior analysis we have found that in the $D\to 4-\epsilon$ limit the general off-shell contribution of photon self-interaction loop to the photon two-point function contains complicated non-vanishing UV and IR divergent terms with existing and new momentum structures, regardless the $\kappa_g$ values we take. To see whether there exists certain remedy to this situation we explore two conditions which have emerged in the prior studies. First we tested the zero mass-shell condition/limit ($p^2\to0$) used in $\theta$-expanded models \cite{Martin:2009vg}. Inspection of Eq's (\ref{FlfinalF1},\ref{FlfinalF2})
and (\ref{PIB1}-\ref{PIB5}) show some simplification but not the full cancelation of the pathological divergences. Such condition clearly appears to be unsatisfactory.

Next we have turned into the other one, namely the special full
rank $\theta^{\mu\nu}$ choice
\begin{equation}
\theta^{\mu\nu}\equiv
\theta^{\mu\nu}_{\sigma_2}=\frac{1}{\Lambda_{\rm NC}^2}
\begin{pmatrix}
0&-1&0&0\\
1&0&0&0\\
0&0&0&-1\\
0&0&1&0
\end{pmatrix}
=\frac{1}{\Lambda_{\rm NC}^2}
\begin{pmatrix}
{i\sigma_2}&0\\
0&{i\sigma_2}
\end{pmatrix}
\equiv\frac{1}{\Lambda_{\rm NC}^2}\:i\sigma_2\otimes I_2,
\label{nondegen}
\end{equation}
with $\sigma_2$ being famous Pauli matrix.

This choice, in 4d Euclidean space\-time\footnote{This condition was used in the renormalizability studies of 4d NCGFT without SW map \cite{Blaschke:2009aw,Blaschke:2010ck}. Note also that this $\theta^{\mu\nu}_{\sigma_2}$ is full rank and thus breaks in general the unitarity if one performs Wick rotation to the Minkowski space\-time.}, induces a relation $(\theta\theta)^{\mu\nu}=-\frac{1}{\Lambda_{\rm NC}^4}g^{\mu\nu}$ similar to the 2d NCGFT discussed in the last subsection.
The tensor structures (\ref{PlD}) then simplifies into two parts
\begin{equation}
\begin{split}
\Pi_{\kappa_g}^{\mu\nu}(p)_4\bigg|^{\theta_{\sigma_2}}
=&\frac{e^2}{(4\pi)^2}\bigg\{\Big[g^{\mu\nu}p^2-p^\mu p^\nu\Big]\, B_I^{\kappa_g}(p)
+(\theta p)^\mu (\theta p)^\nu\, B_{II}^{\kappa_g}(p)\bigg\}
\\&=\frac{e^2}{(4\pi)^2}\bigg\{\Big[g^{\mu\nu}p^2-p^\mu p^\nu\Big]\bigg(B^{\kappa_g}_1+2\frac{B^{\kappa_g}_3}{\Lambda_{\rm NC}^4}-\frac{B^{\kappa_g}_4}{\Lambda_{\rm NC}^8}\bigg)
\\&+(\theta p)^\mu (\theta p)^\nu \bigg(B^{\kappa_g}_2-2\frac{B^{\kappa_g}_5}{\Lambda_{\rm NC}^4}\bigg)\bigg\},
\end{split}
\label{Plfinal4}
\end{equation}
with restored coupling constant $e$ included. Solving $B_I^{\kappa_g}(p)$ and $B_{II}^{\kappa_g}(p)$, and neglecting the IR safe terms, revolves the {\it divergent} parts:
\begin{gather}
B_I^{\kappa_g}(p)\sim\frac{4(\kappa_g-3)^2}{3}\Bigg(\frac{2}{\epsilon}+\ln\Big(\mu^2(\theta p)^2\Big)\Bigg)+\frac{16}{3}\frac{(\kappa_g-3)(\kappa_g+1)}{p^2(\theta p)^2},
\label{BI}\\
B_{II}^{\kappa_g}(p)\sim2\,p^2\frac{(\kappa_g-3)(7\kappa_g-9)}{(\theta p)^2}\Bigg(\frac{2}{\epsilon}+\ln\Big(\mu^2(\theta p)^2\Big)\Bigg)-\frac{16}{3}\frac{(\kappa_g-3)(7\kappa_g-5)}{(\theta p)^4}.
\label{BII}
\end{gather}
We observe immediately that point $\kappa_g=3$ eliminates {\it all} divergences. A careful evaluation of the full photon-loop at this point revolves a simple structures
\begin{equation}
B_I^{\kappa_g=3}(p)
=2\bigg[\frac{56}{3}+{\cal I}\bigg],
\;\;
B_{II}^{\kappa_g=3}(p)
=-\frac{p^2}{(\theta p)^2}\;9\bigg[8-{\cal I}\bigg],
\label{BIBII}
\end{equation}
where 
\begin{equation}
{\cal I}=8\Big({\cal K}\left[0;0,0\right]-6{\cal K}\left[0;1,1\right]\Big)+(\theta p)^2\Big(3{\cal W}\left[1;0,0\right]-16{\cal W}\left[1;1,1\right]\Big)=0,
\label{Izero}
\end{equation}
with $\mathcal{K}\left[\nu;a,b\right]$ and $\mathcal{W}\left[\nu;a,b\right]$ taken from Appendix B. 
The detailed proof of the above vanishing identity $\cal I$ is presented in the Appendix C. Thus, for special choice (\ref{nondegen}) in the $D\to 4-\epsilon$
limit, and  at $\kappa_g=3$ point, we have found
\begin{equation}
B_I^{\kappa_g=3}(p)=\frac{112}{3},\;\;B_{II}^{\kappa_g=3}(p)=-72\frac{p^2}{(\theta p)^2},
\label{BI112BII72}
\end{equation}
as the only one-loop-photon self-interaction corrections to the photon two-point function.

\section{One-loop neutrino self-energy}

\begin{figure}
\begin{center}
\includegraphics[width=12cm,height=5cm]{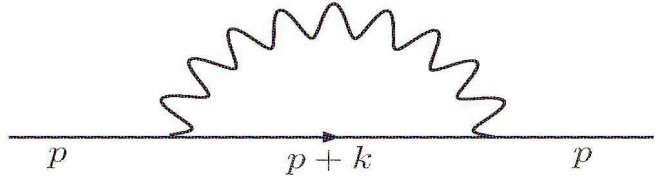}
\end{center}
\caption{Bubble-graph contribution to the neutrino two-point function}
\label{Fig3}
\end{figure}

One-loop contributions as a function of $\kappa_f$
receive the same Lorentz structure as in \cite{Horvat:2011qg}. We now reconfirm that by using the action (\ref{S}) together with the Feynman rule (\ref{Ff}), out of four diagrams in Fig.2 of \cite{Horvat:2011qg}, only the bubble graph (Fig.\ref{Fig3} in this manuscript) gives nonzero contribution.
In the present scenario its contribution reads
\begin{equation}
\Sigma_{\kappa_f}(p)_D=\frac{-1}{(4\pi)^2}\bigg[\gamma_\mu p^\mu\; N_1^{\kappa_f}(p)+\gamma_\mu(\theta\theta p)^\mu\; N^{\kappa_f}_2(p)\bigg].
\label{Sigma1}
\end{equation}
Loop-coefficients $N^{\kappa_f}_{1,2}(p)$, as a functions of an arbitrary dimensions $D$, are given in the Appendix B.

For arbitrary $\kappa_f$ in the limit $D\to 4-\epsilon$ we have obtained the following
loop-coefficients:
\begin{equation}
\begin{split}
N_1^{\kappa_f}(p)=&\;\kappa_f\Bigg\{\bigg[2\kappa_f+\big(\kappa_f-1\big)\bigg(\frac{2}{\epsilon}+\ln{\pi e^{\gamma_E}}+\ln\big(\mu^2(\theta p)^2\big)\bigg)\bigg]
\\&-\frac{p^2(\theta p)^2}{4}\sum\limits_{k=0}^\infty\frac{(p^2(\theta p)^2)^k}{4^kk(k+1)(2k+1)^2(2k+3)\Gamma[2k+4]}
\\&\cdot\bigg[k(k+1)(2k+1)(2k+3)\Big(\kappa_f(2k+3)-1\Big)\Big(\ln\big(p^2(\theta p)^2\big)-2\psi(2k)-\ln4\Big)
\\&+3+28k+46k^2+20k^3-\kappa_f\big(2k+3\big)^2\big(1+8k+8k^2\big)\bigg]
\\&+(\tr\theta\theta)\bigg\{\frac{p^2}{(\theta p)^2}\bigg[\frac{2}{\epsilon}+2+\gamma_E+\ln\pi+\ln\big(\mu^2(\theta p)^2\big)+\frac{8\big(\kappa_f-1\big)}{3\kappa_f(\theta p)^2p^2}\bigg]
\\&-\frac{p^4}{4}\sum\limits_{k=0}^\infty\frac{(p^2(\theta p)^2)^k}{4^kk(k+1)(2k+1)^2(2k+3)}\bigg[k(k+1)(2k+1)(2k+3)
\\&\cdot\Big(\ln\big(p^2(\theta p)^2\big)-2\psi(2k)-\ln4\Big)-2k\Big(14+k(23+10k)\Big)-3\bigg]\bigg\}
\\&+(\theta\theta p)^2\bigg\{2\frac{p^2}{(\theta p)^4}\bigg[\frac{2}{\epsilon}+1+\gamma_E+\ln\pi+\ln\big(\mu^2(\theta p)^2\big)+\frac{16\big(\kappa_f-1\big)}{3\kappa_f(\theta p)^2p^2}\bigg]
\\&+\frac{p^4}{2(\theta p)^2}\sum\limits_{k=0}^\infty\frac{k(p^2(\theta p)^2)^k}{(k+1)(2k+1)^2(2k+3)\Gamma[2k+4]}\bigg[(k+1)(2k+1)(2k+3)
\\&\cdot\Big(\ln\big(p^2(\theta p)^2\big)-2\psi(2k)-\ln4\Big)+16k^2-34k-17\bigg]\bigg\}\Bigg\}\,,
\end{split}
\label{N1}
\end{equation}

\begin{equation}
\begin{split}
N_2^{\kappa_f}(p)=&-\kappa_f\frac{p^2}{(\theta p)^2}\Bigg\{4+\big(\kappa_f-1\big)\bigg[\frac{2}{\epsilon}+\ln{\pi e^{\gamma_E}}+\ln\big(\mu^2(\theta p)^2\big)\bigg]-\frac{16\big(\kappa_f-1\big)}{3(\theta p)^2p^2}
\\&-\frac{p^2(\theta p)^2}{4}\sum\limits_{k=0}^\infty\frac{(p^2(\theta p)^2)^k}{4^kk(k+1)(2k+1)^2(2k+3)\Gamma[2k+4]}
\bigg[k(1+k)(1+2k)(3+2k)
\\&\cdot\Big(1+3k_f+2\big(k_f+1\big)k\Big)\Big(\ln\big(p^2(\theta p)^2\big)-2\psi(2k)-\ln4\Big)-3-9\kappa_f
\\&-4k\bigg(7+21\kappa_f+\Big(20+43\kappa_f
+2k\big(11+4k+4\kappa_f(4+k)\big)\Big)k\bigg)\bigg]\Bigg\}\,.
\end{split}
\label{N2}
\end{equation}
In the expressions for $N_{1,2}^{\kappa_f}(p)$ contributions from both the planar as well as the non-planar graphs are present. For any $\kappa_f\not=1$ our neutrino self energy receive UV, and power as well as logarithmic UV/IR mixing terms. Choosing $\kappa_f=1$ eliminates some of divergences, but not all of them. Imposing the special $\theta^{\mu\nu}_{\sigma_2}$ reduces the contribution to quadratic UV/IR mixing into a
single term from $N_2^{\kappa_f}(p)$, which has two zero points $\kappa_f=0,1$. Only $\kappa_f=0$ can
induce full UV divergence cancelation, by removing the whole $\Sigma_{\kappa_f}(p)_4$.

In the $D\to 2-\epsilon$ limit, we can employ the identities mentioned in Subsection 3.2,
and, after restoring the coupling constant $e$, obtain neutrino self-energy in two dimensions
\begin{equation}
\Sigma_{\kappa_f}(p)_2=-\frac{1}{4\pi}\gamma_\mu p^\mu \;
\Bigg[\frac{-e^2}{p^2}N^{\kappa_f}(p) \Bigg].
\label{SigmaN}
\end{equation}
For arbitrary $\kappa_f$ the above loop-coefficient $N^{\kappa_f}(p)$ has no divergences:
\begin{equation}
\begin{split}
N^{\kappa_f}(p)=&-4\big(\kappa_f^2-2\kappa_f+2\big)-16\big(\kappa_f-1\big)^2f(p,\theta)
\\&=-4\big(\kappa^2_f-2\kappa_f+2\big)
-16\big(\kappa_f-1\big)^2\,p^2\,(\theta p)^2\sum\limits_{k=1}^\infty\frac{(k+1)(p^2)^k((\theta p)^2)^k}{4^k(2k+1)^2(2k+3)^2\Gamma[2k+3]}
\\&\cdot\bigg[(k+1)(2k+1)(2k+3)\Big(\ln\big(p^2(\theta p)^2\big)-2\psi(2k+2)-\ln4\Big)
-\big(4k^2+8k+5\big)\bigg].
\label{Nkappa}
\end{split}
\end{equation}
Here we observe that the commutative limit is safe, there are no UV divergences, IR vanishing terms can be removed by $\kappa_f=1$, only a superficial IR divergence remains for all real $\kappa_f$ values in 
$\Sigma_{\kappa_f}(p)_2$.

\section{Discussion}

In this model we present a $\theta$-exact quantum one-loop contributions to the photon ($\Pi$) and neutrino ($\Sigma$) self-energies and analyze their properties. Our method, an extension of the modified Feynman rule procedure \cite{Filk:1996dm} yields the one-loop quantum corrections for arbitrary dimensions in closed form, as function of the deformation-freedom parameter-space $(\kappa_f,\kappa_g)$, as well as momentum $p^\mu$ and noncommutative parameter $\theta^{\mu\nu}$.
We evaluated the one-loop photon and neutrino self-energies while keeping full parameter-space freedom. Following the extended dimensional regularization technique we expressed the diagrams as $D$-dimensional loop-integrals and identify the relevant momentum structures with corresponding loop-coefficients. We have found that total contribution to photon two-point function satisfies the Ward-(Slavnov-Taylor)  identities for arbitrary dimensions $D$ and for any point ($\kappa_{f}, \kappa_{g}$) in parameter-space:
\begin{equation}
p_\mu \Pi_{(\kappa_f,\kappa_g)}^{\mu\nu}(p)_{D}=p_\mu\left(\Pi_{\kappa_f}^{\mu\nu}(p)_{D}+\Pi_{\kappa_g}^{\mu\nu}(p)_{D}\right)=p_\nu\left(\Pi_{\kappa_f}^{\mu\nu}(p)_{D}+\Pi_{\kappa_g}^{\mu\nu}({p})_{D}\right)=0.
\label{WItotal}
\end{equation}

We observe the following general behavior of one-loop two-point functions in the  $D\to 4-\epsilon$ limit: The total expressions for both the photon and the neutrino self-energy contain the $1/\epsilon$ ultraviolet term, the celebrated UV/IR mixing power terms as well as the logarithmic (soft) UV/IR mixing term. The $1/\epsilon$ divergence is always independent of the noncommutative scale. The logarithmic terms from the $\epsilon$-expansion and the modified Bessel function integral sum into a common term $\ln(\mu^2(\theta p)^2)$, which is divergent both in the IR limit $|p|\to 0$, as well as in the vanishing noncommutativity $\theta \to 0$ limit.

Our evaluation of the four dimensional $\theta$-exact fermion-part contribution to the photon self-energy, i.e. the fermion-loop photon two-point function (\ref{Flfinal})  yields two already known tensor structures \cite{Hayakawa:1999yt,Hayakawa:1999zf,Brandt:2001ud}. The loop-coefficients $F^{\kappa_f}_{1,2}(p)$, on the other hand, exhibit nontrivial $\kappa_f$ dependence. Namely, in the limit $\kappa_f \to 0$ $\Longrightarrow$ $ F^{\kappa_f}_1=F^{\kappa_f}_2=0$, thus the photon self-energy (\ref{Flfinal}) vanishes, while $\kappa_f=1$ appears to be identical to the non SW-map model. Fermion-loop contains UV and logarithmic divergence in $F^{\kappa_f}_1$ for $\kappa_f\neq 0$, while the quadratic UV/IR mixing could be removed by setting $\kappa_f =0,2$ in $F^{\kappa_f}_2(p)$.

The photon-loop contribution to the photon two point function contains various previously unknown new momentum structures with respect to earlier $\theta$-exact results based on $\star$-product only. These higher order in $\theta$ ($\theta\theta\theta\theta$ types) terms suggest certain connection to the open/closed string correspondence \cite{Seiberg:1999vs,Jurco:2001my} (in an inverted way). We consider such connection plausible given the connection between noncommutative field theory and quantum gravity/string theory.

The four dimension expressions of the photon-loop contribution (\ref{PlD}) to self-energy, contains the UV terms, a logarithmic IR singularity as well as quadratic UV/IR mixing terms. This reflects the fact that, up to the $1/\epsilon$ terms, the UV divergence is at most logarithmic, i.e. there is a logarithmic ultraviolet/infrared term representing a soft UV/IR mixing. The results (\ref{PIB1}-\ref{PIB5}) in four dimensions for arbitrary $\kappa_g$ show very complicated structures containing no singular behavior in the infrared ($|p|\to 0$). On the other hand the amplitude $\Pi_{\kappa_g}(p)$ in both, the commutative limit ($\theta\to 0$) and the size-of-the-object limit ($|\theta p|\to 0$), diverge. Also the UV/IR mixings is present for any $\kappa_g$. Inspecting (\ref{PIB1}) to (\ref{PIB5}) together with general structure (\ref{PlD}) we found decouplings of UV and logarithmic IR divergences from the power UV/IR mixing terms. The latter exists in all $B^{\kappa_g}_i$'s.

To simplify the tremendous divergent structures in $B^{\kappa_g}_i$'s at $D\to 4-\epsilon$, we have probed two additional conditions: One which appears to be ineffective is the zero mass-shell condition/limit $p^2\to0$, due to the uncertainty on its own validity when quantum corrections present. The other condition, namely setting $\theta^{\mu\nu}$ to a special full ranked value 
$\theta^{\mu\nu}_{\sigma_2}$ \eqref{nondegen} and working in Euclidean instead of Minkowski space, reduces the number of momentum structures from five to two. Then all divergences and the IR safe contributions desappear at a unique point $(\kappa_f,\kappa_g)=(0,3)$ leaving, in the notation of
Eq. (\ref{WItotal}),
\begin{equation}
\begin{split}
\Pi_{(0,3)}^{\mu\nu}(p)_4\bigg|^{\theta_{\sigma_2}}=
\Pi_{\kappa_g=3}^{\mu\nu}(p)_4\bigg|^{\theta_{\sigma_2}}
=&\frac{e^2p^2}{\pi^2}\Bigg[\frac{7}{3}\bigg(g^{\mu\nu}-\frac{p^\mu p^\nu}{p^2}\bigg)
-\frac{9}{2}\frac{(\theta p)^\mu (\theta p)^\nu}{(\theta p)^2}\Bigg],
\end{split}
\label{Plfinal403}
\end{equation}
as the only one-loop-finite contribution/correction to the photon two-point function.

Considering neutrino self-energy (\ref{Sigma1}), our results extends the prior works \cite{Horvat:2011bs,Horvat:2011qg} by completing the behavior for general $\kappa_f$. Here we discuss some novel behaviors associated with general $\kappa_f$.  The neutrino self-energy does posses power UV/IR mixing phenomenon for arbitrary values of $\kappa_f$, except $\kappa_f=1$. In the limit $\kappa_f\to 0$ all UV, IR divergent terms as well as  constant terms in $N^{\kappa_f}_{1,2}(p)$ vanish; what remains are only the power UV/IR mixing terms. The UV divergence can be localized using the special $\theta$ value \cite{Horvat:2011bs,Horvat:2011qg} in $N^{\kappa_f}_1$ but not in $N^{\kappa_f}_2$. The UV and the power IR divergence in $N^{\kappa_f}_2$ can be removed by setting $\kappa_f=1$.

The general existences of UV/IR mixings for both, photons and neutrinos respectively, in 4d spaces deformed by spacetime noncommutativity at low energies, suggests that the relation of quantum corrections to observations \cite{Horvat:2010km} is not entirely clear.  However, in the context of the UV/IR mixing it is very important to mention a complementary  approach \cite{hep-th/0606248,Abel:2006wj} where NC gauge theories are realized as effective QFT's, underlain by some  more fundamental theory such as string theory. It was claimed that for a large class of more general QFT's above the UV cutoff the phenomenological effects of the UV completion can be quite successfully modeled by a threshold value of the UV cutoff. So, in the presence of a finite UV cutoff no one sort of divergence will ever appear since the problematic phase factors effectively transform the highest energy scale (the UV cutoff) into the lowest one (the IR cutoff). What is more, not only the full scope of noncommutativity is experienced only in the range delimited by the two cutoffs, but for the scale of NC high enough, the whole standard model can be placed below the IR cutoff \cite{Horvat:2010km}. Thus, a way the UV/IR mixing problem becomes hugely less pressing, making a study of the theory at the quantum level much more reliable.

Following the idea of noncommutative two-dimensional gauge theories
\cite{Schwinger:1962tp,Ardalan:2010qb,Armoni:2011pa} we have also studied  the integration dimension $D$ dependence of the loop integrals.
Considering  behavior of divergences in the 2d NCGFT we again present both, the analysis of the photon and the neutrino two-point functions. There exist preferred constraint on, in principle two different, deformation-parameters $(\kappa_f,\kappa_g)$, respectively. That is, the minimal choice $\kappa_f=\kappa_g$ for certain values eliminate all divergences, where $\Pi_{(\kappa_f,\kappa_g)}^{\mu\nu}(p)_2=\Pi_{\kappa_f}^{\mu\nu}(p)_2+\Pi_{\kappa_g}^{\mu\nu}(p)_2$ was computed from (\ref{pkf}) and (\ref{B}), respectively. The neutrino self-energy $\Sigma_{\kappa_f}(p)_2$ was computed from 2d neutrino two-point functions in (\ref{SigmaN}).

The analysis of the fermion-loop contribution to the 2d photon two-point function reveals a result for $g_{\mu\nu}\Pi^{\mu\nu}_{\kappa_f}(p)_2 $ which is finite: $-2e^2/\pi$ or zero, for the freedom parameter $\kappa_f=1$ or $\kappa_f=0,2$, respectively. Cancellation of the modified Bessel function integrals in fermion-loop (\ref{Ff1}) and (\ref{Ff2}) represents in fact the cancellation of non-planar graphs. The nontrivial variation with respect to $\kappa_f$ is however a consequence of summing over additional special function integrals which generalize the non-planar graphs in \cite{Armoni:2011pa}. The photon-loop contribution is also finite and vanishes at $(7\pm\sqrt{21})/2$.


\begin{table}
\begin{center}
\begin{tabular}{ |c|| c| c| }
\hline
$(\kappa_f,\kappa_g)$ & $g_{\mu\nu}\Big(\Pi_{\kappa_f}^{\mu\nu}(p)_2+\Pi_{\kappa_g}^{\mu\nu}(p)_2\Big)$ &
${\fmslash p}\;\Sigma_{\kappa_f}(p)_2$
\\[8pt]
\hline
& & \\
$\kappa_f=\kappa_g=0$   &  $28\frac{\textstyle e^2}{\textstyle \pi}$  & $-2\frac{\textstyle e^2}{\textstyle\pi}\Big(1+2f(p,\theta)\Big)$
\\[8pt]
\hline
& & \\
$\kappa_f=\kappa_g=1$   &  $2\frac{\textstyle e^2}{\textstyle \pi}$ &  $
-\frac{\textstyle  e^2}{\textstyle  \pi}$
\\[8pt]
\hline
& & \\
$\kappa_f=\kappa_g=2$   &  $-12\frac{\textstyle e^2}{\textstyle \pi}$ &  $-2\frac{\textstyle  e^2}{\textstyle  \pi}\Big(1+2f(p,\theta)\Big)$
\\[8pt]
\hline
& & \\
$\kappa_f=\kappa_g=3$   &  $-14\frac{\textstyle  e^2}{\textstyle \pi}$ &  $-\frac{\textstyle e^2}{\textstyle\pi}\Big(5+16f(p,\theta)\Big)$
\\[8pt]
\hline
\end{tabular}
\end{center}
\caption{Photon ($\Pi$) and neutrino ($\Sigma$) self-energies in 2d NCGFT given for typical values of the deformation parameters, in the simplest case $\kappa_f=\kappa_g$. The function $f(p,\theta)$ could be deduced from 2d neutrino two-point functions. }
\end{table}

The behavior of the 2d neutrino two-point function (\ref{SigmaN}) shows a divergent behaviour in the infrared ($|p|\to 0$) for arbitrary $\kappa_f$. However, for arbitrary $\kappa_f$ the commutative limit ($\theta\to 0$) is smooth. All above 2d NCGFT properties are summarized in Table 1.

\section{Conclusion}
After having defined and explained the full noncommutative action-model origin of the deformation parameters $\kappa_f, \kappa_g$, we obtained the relevant Feynman rules. The one-loop photon self-energy in four dimensions contains the UV divergence and UV/IR mixing terms dependent on the freedom parameters $\kappa_f$ and $\kappa_g$. The introduction of the freedom parameters univocally has a  potential to improve the situation regarding cancellation of divergences, since certain choices for $\kappa_f$ and $\kappa_g$ could make some of the terms containing singularities to vanish. In conclusion, our main result in four-dimensional space is that we have under full control all pathological terms as a consequence of the introduction of the deformation-freedom parameter-space $(\kappa_f,\kappa_g)$ and a special choice for $\theta^{\mu\nu}$. In particular, working in the 4d Euclidean space with a special full rank of $\theta^{\mu\nu}_{\sigma_2}$ and setting $(\kappa_f,\kappa_g)=(0,3)$, the fermion plus the photon-loop contribution to $\Pi^{\mu\nu}_{(\kappa_f,\kappa_g)}(p)_4$ contain only two finite terms, i.e. all divergent terms are eliminated. In this case the neutrino two-point function vanishes. In two-dimensional space, the photon self-energy is finite, while the neutrino self-energy still contains an IR divergence, for any choice of the deformation-freedom parameters $\kappa_f, \kappa_g$. From Table 1 we see  that in the minimal 2d NC $\theta$-exact U(1) action with $\kappa_f=\kappa_g=\big[0;2\big]$, the photon self-energy is finite $\big[+28e^2/2\pi; -12e^2/\pi\big]$ (with the opposite sign), while the neutrino self-energy becomes the same for both choices. Due to the dominance of the photon-loop contributions the overall sign for the 2d photon self-energy is changed and for some values of $\kappa_f=\kappa_g$ its contribution  gets enhanced by a large factor with respect to the previous result \cite{Armoni:2011pa}.  All four choices for the deformation-freedom parameters in Table 1 produce no divergences at all, up to the artificial IR divergence for the neutrino self energy. The above profound structure in 2d NCGFT suggests further study in the gauge/gravity duality framework \cite{Erdmenger:2012zz,Hashimoto:1999ut,Maldacena:2000vw,Landsteiner:2007bd}, with the possibility of the important  connection of  the 3d NCGFT with the 3d gravity,
in particular.

\section{Acknowledgment}
J.T. would like to acknowledge support of
Max-Planck-Institute for Physics, Munich, for hospitality,
and J. Erdmenger and W. Hollik for fruitful discussions. Special thanks goes to D. Blaschke for pointing out to us the special choice of the noncommutative parameter. We would like to thank
C.P. Martin and P. Schupp for various helpful comments on the manuscript. A great deal of computation was done by using ${\rm Mathematica}$~8.0 \cite{mathematica} plus tensor algebra package xAct \cite{xAct}. The work of R.H., J.T. and J.Y. are supported by the Croatian Ministry of Science, Education and Sports under Contract Nos. 098-0982930-2872. The work of A.I. is supported by the Croatian Ministry of Science, Education and Sports under Contracts Nos. 119-0982930-1016.

\appendix

\section{Integral parametrization}

In this part we illustrate the way we have performed the computation of the integrals
which differ from regular ones by the existence of a non-quadratic
$k\theta p$ denominators. The key point was to introduce
the HQET parametrization \cite{Grozin:2000cm}, represented as follows
\begin{equation}
\frac{1}{a_1^{n_1} a_2^{n_2}}=
 \frac{\Gamma(n_1+n_2)}{\Gamma(n_1)\Gamma(n_2)}
 \int_0^\infty\frac{i^{n_1}y^{n_1-1} dy}{(ia_1y + a_2)^{n_1+n_2}}\,.
\label{Grozin}
\end{equation}

To perform computations of our integrals, we first use
the Feynman parametrization on the quadratic denominators,
then the HQET parametrization help us to combine
the quadratic and linear denominators. For example
\begin{eqnarray}
\frac{1}{k^2(p+k)^2}\frac{1}{k\theta p}&=&2i\int\limits_0^1
dx\int\limits_0^\infty dy
\Big[(k^2+i\epsilon)(1-x)+\big((p+k)^2+i\epsilon\big)x+iy(k\theta p)\Big]^{-3}.
\label{denom456}
\end{eqnarray}
After employing the Schwinger parametrization, the phase factors from (\ref{Ff}) can be absorbed by redefining the $y$ integral. This way we obtain
\begin{eqnarray}
\frac{2-e^{ik\theta p}-e^{-ik\theta p}}{k^2(p+k)^2(k\theta p)}\cdot\{\rm numerator\}
&=&
2i\int\limits_0^1\,dx\int\limits_0^{\frac{1}{\lambda}}dy\int\limits_0^\infty\,d\lambda\lambda^2
e^{-\lambda\big(l^2+x(1-x)p^2+\frac{y^2}{4}(\theta p)^2\big)}
\nonumber\\
&\cdot&\{y\;\rm odd\;terms\;of\;the\;numerator\},
\label{denomnum}
\end{eqnarray}
with loop-momenta being $l=k+xp+\frac{i}{2}y(\theta p)$. By this means the $y$-integral
limits take the places of planar/nonplanar parts of the loop integral. For higher negative
power(s) of $k\theta p$, the parametrization follows the same way except the appearance of the additional $y$-integrals which lead to {\em finite} hypergeometric functions \cite{regularizedgeneralizedhypergeometric}.

\section{Loop Integrals}
Employing the aforementioned parametrization we observe that all loop integrals we have computed can be expressed using two series of integrals in addition to the usual planar dimensional regularization formulas. These integrals, denoted as $\mathcal{K}$ and $\mathcal{W}$, are defined as follows:
\begin{gather}
\mathcal{K}[\nu;a,b]=2^{-\nu}(\theta p)^\nu\int\limits_0^1\,dx\,x^a(1-x)^b  X^{-\nu}  K_\nu[X]
\\
\mathcal{W}[\nu;a,b]=\int\limits_0^1\,dx\,x^a(1-x)^b W_\nu[X]
\end{gather}
where $K_\nu[X]$ is the modified Bessel function of second kind, while
\begin{eqnarray}
W_\nu[X]&=&(\theta p)^{-2\nu}\Bigg(X^{2\nu}\Gamma\left[-\nu\right] {}_1F_2\left[\frac{1}{2};\frac{3}{2},\nu+1;\frac{X^2}{4}\right]
\nonumber\\
&-&\frac{2^{2\nu}}{1-2\nu}\Gamma\left[\nu\right]{}_1F_2\left[\frac{1-2\nu}{2};1-\nu,\frac{3-2\nu}{2};\frac{X^2}{4}\right]\Bigg).
\end{eqnarray}
The variable $X$ is defined in \eqref{X}.

Loop coefficients $F_i^{\kappa_f}(p)$ involves integral $\mathcal{K}$'s only:
\begin{gather}
\begin{split}
F_1^{\kappa_f}(p) =
 -4{\rm Dim}(Cl[[d]])(4\pi)^{2-\frac{D}{2}}\mu^{d-D}
&\kappa_f^2\Bigg(\Gamma\Big(2-\frac{D}{2}\Big) \frac{\left(\Gamma(\frac{D}{2})\right)^2}{\Gamma(D)} (p^2)^{\frac{D}{2}-2}
-2\mathcal{K}\left[\frac{D}{2}-2;1,1\right]\Bigg),
\end{split}
\label{F1}\\
\begin{split}
F_{2}^{\kappa_f}(p)
 ={\rm Dim}(Cl[[d]])(4\pi)^{2-\frac{D}{2}}\mu^{d-D}
 &\kappa_f\Bigg(\big(\kappa_f-1\big) \Big(\frac{4}{(\theta p)^2}\Big)^{\frac{D}{2}} \frac{2\Gamma(\frac{D}{2})}{D-1}
 -2\kappa_f \;\mathcal{K}\left[\frac{D}{2};0,0\right]\Bigg),
\end{split}
\label{F2}
\end{gather}
while the loop coefficients $B^{\kappa_g}_i(p)$'s and $N^{\kappa_f}_i(p)$'s contain both integrals, the $\mathcal{K}$'s and the $\mathcal{W}$'s, respectively:
\begin{equation}
\begin{split}
B^{\kappa_g}_1(p)=&(4\pi)^{2-\frac{D}{2}}\mu^{d-D}\Bigg\{-\frac{2^{2-D}\pi^{\frac{3}{2}}\csc\frac{D\pi}{2}(p^2)^{\frac{D}{2}-2}}{\Gamma\left(\frac{D+1}{2}\right)}
\\&\cdot
\bigg\{D^2(\kappa_g-3)^2-D\Big(\kappa_g(3\kappa_g-22)+37\Big)
\\&-2\Big(\kappa_g(\kappa_g+2)
-11\Big)
+\Big((D-2)\kappa_g^2+2D\kappa_g+3D-4\Big)(\tr\theta\theta)\frac{p^2}{(\theta p)^2}
\\&+2\Big((D-2)\kappa_g^2+2D\kappa_g+(D-2)\Big)(\theta\theta p)^2\frac{p^2}{(\theta p)^4}\bigg\}
\\& -8(\kappa_g-2)^2\mathcal{K}\left[\frac{D}{2}-2;0,0\right]
\\&+8\Big(D(\kappa_g-3)^2+3(\kappa_g^2-2\kappa_g-1)\Big)\mathcal{K}\left[\frac{D}{2}-2;1,1\right]
\\&+\Big(-2\kappa_g^2+8\kappa_g+D-11\Big)(\theta p)^2\mathcal{W}\left[\frac{D}{2}-1;0,0\right]
\\&+2\Big(3(\kappa_g-1)^2+D(\kappa_g^2-6\kappa_g+7)\Big)(\theta p)^2\mathcal{W}\left[\frac{D}{2}-1;1,1\right]
\\&+(\tr\theta\theta)\frac{p^2}{(\theta p)^2}\bigg\{4(\kappa_g+2)^2\mathcal{K}\left[\frac{D}{2}-2;0,0\right]
\\&-\frac{8(D+1)(\kappa_g-1)^2}{D-1}\mathcal{K}\left[\frac{D}{2}-2;1,1\right]+\frac{8(D+1)(\kappa_g-1)^2}{D-1}p^{-2}\mathcal{K}\left[\frac{D}{2}-1;0,0\right]
\\&+(\kappa_g^2+2)(\theta p)^2\mathcal{W}\left[\frac{D}{2}-1;0,0\right]-\frac{2(D+1)(\kappa_g-1)^2}{D-1}(\theta p)^2\mathcal{W}\left[\frac{D}{2}-1;1,1\right]\bigg\}
\\&+(\theta\theta p)^2\frac{p^2}{(\theta p)^4}\bigg\{8(\kappa_g^2+1)\mathcal{K}\left[\frac{D}{2}-2;0,0\right]
-\frac{16D(\kappa_g-1)^2}{D-1}\mathcal{K}\left[\frac{D}{2}-2;1,1\right]
\\&+\frac{8D(\kappa_g-1)^2}{D-1}p^{-2}\mathcal{K}\left[\frac{D}{2}-1;0,0\right]
+(\kappa_g^2+1)(\theta p)^2\mathcal{W}\left[\frac{D}{2}-1;0,0\right]
\\&+\frac{2(D+1)(\kappa_g-1)^2}{D-1}(\theta p)^2\mathcal{W}\left[\frac{D}{2}-1;1,1\right]\bigg\}\Bigg\},
\end{split}
\label{PIntB1}
\end{equation}
\begin{equation}
\begin{split}
B^{\kappa_g}_2(p)=&(4\pi)^{2-\frac{D}{2}}\mu^{d-D}p^2(\theta p)^{-2}\Bigg\{\frac{2^{2-D}\pi^{\frac{3}{2}}\csc\frac{D\pi}{2}(p^2)^{\frac{D}{2}-2}}{\Gamma\left(\frac{D+1}{2}\right)}
\\&\cdot
\bigg\{-2D^3(\kappa_g-1)^2+D^2\Big(\kappa_g(7\kappa_g-6)-5\Big)
\\&+D\Big(19-\kappa_g(9\kappa_g-2)\Big)+2\Big(\kappa_g(\kappa_g+2)-7\Big)
\\&-2\Big(3-2D+\kappa_g(\kappa_g-2)\Big)(\tr\theta\theta)\frac{p^2}{(\theta p)^2}
-8(\kappa_g-1)^2(\theta\theta p)^2\frac{p^2}{(\theta p)^4}\bigg\}
\\&+8\Big((D-3)\kappa_g^2+(12-4D)\kappa_g+3D-8\Big)\mathcal{K}\left[\frac{D}{2}-2;0,0\right]
\\&+8\Big(-2D^3(\kappa_g-1)^2+D^2(\kappa_g^2+14\kappa_g-17)+D(7\kappa_g^2-46\kappa_g+37)
\\&-12\kappa_g^2+44\kappa_g-42\Big)
\frac{1}{D-1}\mathcal{K}\left[\frac{D}{2}-2;1,1\right]
\\&+
4\Big(\kappa_g^2(D-2)(D-7)-2\kappa_g(D-2)(D-9)+2D^3-9D^2-7D+38\Big)
\\&\cdot
\frac{1}{p^2(D-1)}\mathcal{K}\left[\frac{D}{2}-1;0,0\right]
\\&+\Big(2(D+1)\kappa_g^2-2(4D+2)\kappa_g+8D+13\Big)(\theta p)^2\mathcal{W}\left[\frac{D}{2}-1;0,0\right]
\\&-2\Big(2D^3(\kappa_g-1)^2+D^2(\kappa_g^2-22\kappa_g+25)+2D(\kappa_g^2+6\kappa_g-2)+(\kappa_g-1)^2\Big)
\\&\cdot\frac{(\theta p)^2}{D-1}\mathcal{W}\left[\frac{D}{2}-1;1,1\right]
\\&+(\tr\theta\theta)\frac{p^2}{(\theta p)^2}\bigg\{8 \mathcal{K}\left[\frac{D}{2}-2;0,0\right]-\frac{8(\kappa_g-1)^2}{D-1} \mathcal{K}\left[\frac{D}{2}-2;1,1\right]
\\&+\frac{4D(\kappa_g-1)^2}{D-1}p^{-2}\mathcal{K}\left[\frac{D}{2}-1;0,0\right]
\\&+(\theta p)^2\mathcal{W}\left[\frac{D}{2}-1;0,0\right]-\frac{2(\kappa_g-1)^2}{D-1} (\theta p)^2\mathcal{W}\left[\frac{D}{2}-1;1,1\right]\bigg\}
\\&+(\theta\theta p)^2\frac{p^2}{(\theta p)^4}(\kappa_g-1)^2\bigg\{\frac{8(D-4)}{D-1}\mathcal{K}\left[\frac{D}{2}-2;1,1\right]
\\&+\frac{4D(D+2)}{D-1}p^{-2}\mathcal{K}\left[\frac{D}{2}-1;0,0\right]
-\frac{6}{D-1}(\theta p)^2\mathcal{W}\left[\frac{D}{2}-1;1,1\right]\bigg\}\Bigg\},
\end{split}
\label{PIntB2}
\end{equation}
\begin{equation}
\begin{split}
B^{\kappa_g}_3(p)=&(4\pi)^{2-\frac{D}{2}}\mu^{d-D}p^4(\theta p)^{-2}\Bigg\{\frac{2^{2-D}\pi^{3/2}\csc\frac{D\pi}{2}(p^2)^{\frac{D}{2}-2}}{\Gamma\left(\frac{D+1}{2}\right)}
\\&\cdot\bigg(1-3\kappa_g(\kappa_g-2)+D(\kappa_g+1)(\kappa_g-3)\bigg)
\\&-8\mathcal{K}\left[\frac{D}{2}-2;0,0\right]-8\frac{(D-3)\kappa_g^2-2(D-5)\kappa_g-D-3}{D-1}\mathcal{K}\left[\frac{D}{2}-2;1,1\right]
\\&+4\frac{(D-3)\kappa_g^2+(14-6D)\kappa_g+2D^2-3D-3}{D-1}p^{-2}\mathcal{K}\left[\frac{D}{2}-1;0,0\right]
\\&-(4\kappa_g-D+2)(\theta p)^2\mathcal{W}\left[\frac{D}{2}-1;0,0\right]
\\&+4\Big(\kappa_g^2-2(D+1)\kappa_g+D^2+1\Big)\frac{(\theta p)^2}{D-1}\mathcal{W}\left[\frac{D}{2}-1;1,1\right]\Bigg\},
\end{split}
\label{PIntB3}
\end{equation}
\begin{equation}
\begin{split}
B^{\kappa_g}_4(p)=&(4\pi)^{2-\frac{D}{2}}\mu^{d-D}p^4(\theta p)^{-4}\Bigg\{\frac{2^{3-D}\pi^{3/2}\csc\frac{D\pi}{2}(p^2)^{\frac{D}{2}-2}}{\Gamma\left(\frac{D+1}{2}\right)}\bigg((D-1)(\kappa_g+1)^2\bigg)
\\&+\Big(8(1-D)+32\kappa_g\Big)\mathcal{K}\left[\frac{D}{2}-2;0,0\right]
\\&-\frac{8(2D-1)(\kappa_g^2-6\kappa_g+2D-1)}{D-1}\mathcal{K}\left[\frac{D}{2}-2;1,1\right]
\\&+\frac{4D(\kappa_g^2-6\kappa_g+2D-1)}{D-1}p^{-2}\mathcal{K}\left[\frac{D}{2}-1;0,0\right]
\\&+(4\kappa_g+1-D)(\theta p)^2\mathcal{W}\left[\frac{D}{2}-1;0,0\right]
\\&-\frac{2D(\kappa_g^2-6\kappa_g+2D-1)}{D-1}(\theta p)^2\mathcal{W}\left[\frac{D}{2}-1;1,1\right]\Bigg\},
\end{split}
\label{PIntB4}
\end{equation}
\begin{equation}
\begin{split}
B^{\kappa_g}_5(p)=&(4\pi)^{2-\frac{D}{2}}\mu^{d-D}p^4(\theta p)^{-4}\Bigg\{\frac{2^{4-D}\pi^{\frac{3}{2}}\csc\frac{D\pi}{2}(p^2)^{\frac{D}{2}-2}}{\Gamma\left(\frac{D+1}{2}\right)}\bigg(\kappa_g^2+(D-3)\kappa_g+D\bigg)
\\&+16\kappa_g\mathcal{K}\left[\frac{D}{2}-2;0,0\right]+\frac{16(2D-\kappa_g)(\kappa_g-1)}{D-1} \mathcal{K}\left[\frac{D}{2}-2;1,1\right]
\\&+\frac{8D(\kappa_g-1)(\kappa_g-2)}{D-1}p^{-2}\mathcal{K}\left[\frac{D}{2}-1;0,0\right]+2\kappa_g(\theta p)^2\mathcal{W}\left[\frac{D}{2}-1;0,0\right]
\\&+\frac{(D+1-\kappa_g)(\kappa-1)}{D-1}(\theta p)^2\mathcal{W}\left[\frac{D}{2}-1;1,1\right]\Bigg\},
\end{split}
\label{PIntB5}
\end{equation}
\begin{equation}
\begin{split}
N^{\kappa_f}_1(p)=&(4\pi)^{2-\frac{D}{2}}\mu^{d-D}\Bigg\{\frac{\pi^{\frac{3}{2}}2^{3-D}\csc\frac{D\pi}{2}p^{D-4}}{\Gamma\left[\frac{D-1}{2}\right]}
\\&\cdot\bigg\{(D-3)(\kappa_f-1)-(\tr\theta\theta)\frac{p^2}{(\theta p)^2}-2(\theta\theta p)^2\frac{p^2}{(\theta p)^4}\bigg\}
\\&-4(\kappa_f-1)\Big((2D-3)\kappa_f-D\Big)\mathcal{K}\left[\frac{D}{2}-2;1,0\right]
\\&+8(D-1)(\kappa_f-1)^2\mathcal{K}\left[\frac{D}{2}-2;2,0\right]
-(D-3)\kappa_f^2(\theta p)^2\mathcal{W}\left[\frac{D}{2}-1;0,0\right]
\\&-\Big((D+1)\kappa_f^2+(1-3D)\kappa_f+1+D\Big)(\theta p)^2\mathcal{W}\left[\frac{D}{2}-1;1,0\right]
\\&+2D(\kappa_f-1)^2(\theta p)^2\mathcal{W}\left[\frac{D}{2}-1;2,0\right]-(\tr\theta\theta)\frac{p^2}{(\theta p)^2}\bigg\{4\mathcal{K}\left[\frac{D}{2}-2;0,1\right]
\\&+\frac{8D(\kappa_f-1)}{D-1}\mathcal{K}\left[\frac{D}{2}-2;1,1\right]+\frac{4(\kappa_f-1)}{D-1}p^{-2}\mathcal{K}\left[\frac{D}{2}-1;0,0\right]
\\&-(\theta p)^2\mathcal{W}\left[\frac{D}{2}-1;0,1\right]
+\frac{2D(\kappa_f-1)}{D-1}(\theta p)^2\mathcal{W}\left[\frac{D}{2}-1;1,1\right]\bigg\}
\\&+(\theta\theta p)^2\frac{p^2}{(\theta p)^4}\bigg\{-8\mathcal{K}\left[\frac{D}{2}-2;0,1\right]-\frac{8(2D-1)(\kappa_f-1)}{D-1}\mathcal{K}\left[\frac{D}{2}-2;1,1\right]
\\&+\frac{4D(\kappa_f-1)}{D-1}p^{-2}\mathcal{K}\left[\frac{D}{2}-1;0,0\right]-(\theta p)^2\mathcal{W}\left[\frac{D}{2}-1;0,1\right]
\\&-\frac{2D(\kappa_f-1)}{D-1}(\theta p)^2\mathcal{W}\left[\frac{D}{2}-1;1,1\right]\bigg\}\Bigg\},
\end{split}
\end{equation}
\begin{equation}
\begin{split}
N^{\kappa_f}_2(p)=&(4\pi)^{2-\frac{D}{2}}\mu^{d-D}\Bigg\{-2\kappa_f(\kappa_f-1)\rm{B}\left[\frac{D}{2},\frac{D}{2}-1\right]\Gamma\left[2-\frac{D}{2}\right]p^{D-4}\frac{p^2}{(\theta p)^2}
\\&-\kappa_f(\kappa_f-1)\bigg\{\frac{4(D+1)}{D-1}\mathcal{K}\left[\frac{D}{2}-2;1,0\right]+\frac{8}{D-1}\mathcal{K}\left[\frac{D}{2}-2;2,0\right]
\\&-\frac{4(D-2)}{D-1}p^{-2}\mathcal{K}\left[\frac{D}{2}-1;0,0\right]\bigg\}
\\&-\kappa_f(\theta p)^2\bigg\{-2\mathcal{W}\left[\frac{D}{2}-1;0,0\right]+\frac{(1-3D)\kappa_f+5D-3}{D-1}\mathcal{W}\left[\frac{D}{2}-1;1,0\right]
\\&+\frac{2D(\kappa_f-1)}{D-1}\mathcal{W}\left[\frac{D}{2}-1;2,0\right]\bigg\}\Bigg\}.
\end{split}
\end{equation}

\section{Proof for the vanishing identity $\mathcal{I}$}

In this section we evaluate the equation \eqref{Izero} and verify that the integral identity $\mathcal{I}$ equals to zero. First we compute each of the special function integrals
\begin{equation}
\begin{split}
\mathcal{K}\left[0;0,0\right]=&\int\limits_0^1\,dx\,K_0[X]
\\
=&\int\limits_0^1\;(-)\sum\limits_{k=0}^\infty\frac{x^{k}(1-x)^{k}}{\left(\Gamma\left[k+1\right]\right)^2}\left(\frac{p^2(\theta p)^2}{4}\right)^k \left(\frac{1}{2}\ln\frac{x(1-x)p^2(\theta p)^2}{4}-\psi(k+1)\right)
\\
=&-\sum\limits_{k=0}^\infty\frac{1}{\Gamma\left[2k+2\right]}\left(\frac{p^2(\theta p)^2}{4}\right)^k\left(\frac{1}{2}\ln\frac{p^2(\theta p)^2}{4}-\psi(2k+2)\right),
\end{split}
\end{equation}

\begin{equation}
\begin{split}
\mathcal{K}\left[0;1,1\right]=&\int\limits_0^1\,dx\;x(1-x)K_0[X]
\\
=&\int\limits_0^1\;(-)\sum\limits_{k=0}^\infty\frac{x^{k+1}(1-x)^{k+1}}{\left(\Gamma\left[k+1\right]\right)^2}\left(\frac{p^2(\theta p)^2}{4}\right)^k \left(\frac{1}{2}\ln\frac{x(1-x)p^2(\theta p)^2}{4}-\psi(k+1)\right)
\\
=&-\sum\limits_{k=0}^\infty\frac{(k+1)^2}{\Gamma\left[2k+4\right]}\left(\frac{p^2(\theta p)^2}{4}\right)^k\left(\frac{1}{2}\ln\frac{p^2(\theta p)^2}{4}+\frac{1}{k+1}-\psi(2k+4)\right),
\end{split}
\end{equation}

\begin{equation}
\begin{split}
\mathcal{W}\left[1;0,0\right]=&\int\limits_0^1\,dx\,W_1[X]
\\
=&\int\limits_0^1\;-4(\theta p)^{-2}+4(\theta p)^{-2}\sum\limits_{k=0}^\infty\frac{2x^k(1-x)^k}{\Gamma[k+1]\Gamma[k+2](2k+1)}\left(\frac{p^2(\theta p)^2}{4}\right)^{k+1}
\\
\cdot&\left(\frac{1}{2}\ln\frac{x(1-x)p^2(\theta p)^2}{4}+\frac{1}{2}\psi(k+1)+\frac{1}{2}\psi(k+2)+\frac{1}{2k+1}\right)
\\=&-4(\theta p)^{-2}+4(\theta p)^{-2}\sum\limits_{k=0}^\infty\frac{k+1}{\Gamma[2k+4](2k+1)}\left(\frac{p^2(\theta p)^2}{4}\right)^{k+1}\bigg(\ln\frac{p^2(\theta p)^2}{4}
\\
&+\frac{1}{k+1}-\frac{2}{2k+1}-2\psi(2k+4)\bigg),
\end{split}
\end{equation}
and
\begin{equation}
\begin{split}
\mathcal{W}\left[1;1,1\right]=&\int\limits_0^1\,dx\;x(1-x)W_1[X]
\\
=&\int\limits_0^1\,-4x(1-x)(\theta p)^{-2}+4(\theta p)^{-2}\sum\limits_{k=0}^\infty\frac{2x^{k+1}(1-x)^{k+1}}{\Gamma[k+1]\Gamma[k+2](2k+1)}
\\
\cdot&\left(\frac{p^2(\theta p)^2}{4}\right)^{k+1}\left(\frac{1}{2}\ln\frac{x(1-x)p^2(\theta p)^2}{4}+\frac{1}{2}\psi(k+1)+\frac{1}{2}\psi(k+2)-\frac{1}{2k+1}\right)
\\&=-\frac{2}{3}(\theta p)^{-2}+(\theta p)^{-2}\sum\limits_{k=0}^\infty\frac{4(k+1)(k+2)^2}{\Gamma\left[2k+6\right](2k+1)}
\\&\cdot\left(\frac{p^2(\theta p)^2}{4}\right)^{k+1}\bigg(\ln\frac{p^2(\theta p)^2}{4}
+\frac{1}{k+1}+\frac{2}{k+2}-\frac{2}{2k+1}-2\psi(2k+6)\bigg).
\end{split}
\end{equation}
Here $\psi(z)$ is the polygamma function $\psi(z)=d_z\ln\Gamma[z]$, which satisfies the recurrence relation
$\psi(z+1)=\psi(z)+z^{-1}$. Now we have
\begin{equation}
\begin{split}
\mathcal{I}=&-8\bigg[\left(\frac{1}{2}\ln\frac{p^2(\theta p)^2}{4}-\psi(2)\right)-\frac{6}{\Gamma[4]}\left(\frac{1}{2}\ln\frac{p^2(\theta p)^2}{4}+1-\psi(4)\right)\bigg]-\bigg(3\cdot4
\\&-16\cdot\frac{2}{3}\bigg)+\sum\limits_{k=0}^\infty\left(\frac{p^2(\theta p)^2}{4}\right)^{k+1}\Bigg[\frac{-8}{\Gamma\left[2k+4\right]}\bigg(\frac{1}{2}\ln\frac{p^2(\theta p)^2}{4}-\psi(2k+4)\bigg)
\\&+8\cdot6\frac{(k+2)^2}{\Gamma\left[2k+6\right]}\bigg(\frac{1}{2}\ln\frac{p^2(\theta p)^2}{4}
+\frac{1}{k+2}-\psi(2k+6)\bigg)+\frac{3\cdot4(k+1)}{\Gamma[2k+4](2k+1)}
\\&\cdot\bigg(\ln\frac{p^2(\theta p)^2}{4}+\frac{1}{k+1}-\frac{2}{2k+1}-2\psi(2k+4)\bigg)-\frac{16\cdot4(k+1)(k+2)^2}{\Gamma\left[2k+6\right](2k+1)}
\\&\cdot\bigg(\ln\frac{p^2(\theta p)^2}{4}+\frac{1}{k+1}+\frac{2}{k+2}-\frac{2}{2k+1}-2\psi(2k+6)\bigg)\Bigg]
\\=&\sum\limits_{k=0}^\infty\left(\frac{p^2(\theta p)^2}{4}\right)^{k+1}\Bigg[\bigg(\ln\frac{p^2(\theta p)^2}{4}-2\psi(2k+4)\bigg)\bigg(-\frac{4}{\Gamma\left[2k+4\right]}+\frac{24(k+2)^2}{\Gamma[2k+6]}
\\&+\frac{12(k+1)}{\Gamma\left[2k+4\right](2k+1)}-\frac{64(k+1)(k+2)^2}{\Gamma\left[2k+6\right](2k+1)}\bigg)+\frac{12(k+1)}{\Gamma[2k+4](2k+1)}
\\&\cdot\bigg(\frac{1}{k+1}-\frac{2}{2k+1}\bigg)-\frac{64(k+1)(k+2)^2}{\Gamma\left[2k+6\right](2k+1)}\bigg(\frac{1}{k+1}+\frac{1}{k+2}-\frac{2}{2k+1}
\\&-\frac{2}{2k+5}\bigg)+\frac{48(k+2)^2}{\Gamma[2k+6]}\bigg(\frac{1}{2k+4}-\frac{1}{2k+5}\bigg)\Bigg].
\end{split}
\end{equation}
One can then see that
\begin{equation}
\begin{split}
&-\frac{4}{\Gamma\left[2k+4\right]}+\frac{24(k+2)^2}{\Gamma[2k+6]}
+\frac{12(k+1)}{\Gamma\left[2k+4\right](2k+1)}-\frac{64(k+1)(k+2)^2}{\Gamma\left[2k+6\right](2k+1)}
\\&=\frac{4}{\Gamma\left[2k+4\right]}\bigg(-1+\frac{3(k+2)}{2k+5}+\frac{3(k+1)}{2k+1}-\frac{8(k+1)(k+2)}{(2k+1)(2k+5)}\bigg)=0,
\end{split}
\end{equation}
\begin{equation}
\begin{split}
&\frac{12(k+1)}{\Gamma[2k+4](2k+1)}\bigg(\frac{1}{k+1}-\frac{2}{2k+1}\bigg)-\frac{64(k+1)(k+2)^2}{\Gamma\left[2k+6\right](2k+1)}
\\&\cdot\bigg(\frac{1}{(k+2)(2k+5)}-\frac{1}{(k+1)(2k+1)}\bigg)+\frac{48(k+2)^2}{\Gamma[2k+6]}\bigg(\frac{1}{2k+4}-\frac{1}{2k+5}\bigg)
\\&=\frac{1}{\Gamma\left[2k+4\right]}\bigg[\frac{-12}{(2k+1)^2}+\frac{96(2k+3)}{(2k+1)^2(2k+5)^2}+\frac{12}{(2k+5)^2}\bigg]=0,
\end{split}
\end{equation}
thus
\begin{equation}
\mathcal{I}=0.
\end{equation}

\end{document}